\documentclass[preprint,prX]{revtex4}
\usepackage{graphicx}% Include figure filesg
\usepackage{bm}% bold math

\begin{document}

\title{An Upper Limit on the Stochastic Gravitational-Wave Background
of Cosmological Origin}

\author{B.~P.~Abbott$^{1}$,
R.~Abbott$^{1}$,
F. Acernese$^{2ac}$,
R.~Adhikari$^{1}$,
P.~Ajith$^{3}$,
B.~Allen$^{3,4}$,
G.~Allen$^{5}$,
M. Alshourbagy$^{6ab}$,
R.~S.~Amin$^{7}$,
S.~B.~Anderson$^{1}$
% ...
}
\address{$^{1,*}$LIGO - California Institute of Technology, Pasadena, CA  91125, USA }
\address{$^{2,\dagger}$INFN, sezione di Napoli $^a$; Universit\`a di Napoli 'Federico II'$^b$ Complesso Universitario di Monte S.Angelo, I-80126 Napoli; Universit\`a di Salerno, Fisciano, I-84084 Salerno$^c$, Italy}
\address{$^{3,*}$Albert-Einstein-Institut, Max-Planck-Institut f\"ur Gravitationsphysik, D-30167 Hannover, Germany}
\address{$^{4,*}$University of Wisconsin-Milwaukee, Milwaukee, WI  53201, USA }
\address{$^{5,*}$Stanford University, Stanford, CA  94305, USA }
\address{$^{6,\dagger}$INFN, Sezione di Pisa$^a$; Universit\`a di Pisa$^b$; I-56127 Pisa; Universit\`a di Siena, I-53100 Siena$^c$, Italy}
\address{$^{7,*}$Louisiana State University, Baton Rouge, LA  70803, USA }
% ...
\address{$^{*}$The LIGO Scientific Collaboration }
\address{$^{\dagger}$The Virgo Collaboration }

%\author{The LIGO Scientific Collaboration and the Virgo Collaboration}
%\footnote{Please see the online supplement for the individual author lists of the two collaborations.}
%\date{\today}

\maketitle

{\bf A stochastic background of gravitational waves is expected to arise
from a superposition of a large number of unresolved gravitational-wave
sources of astrophysical and cosmological origin. It is
expected to carry unique signatures from the earliest epochs in the
evolution of the universe, inaccessible to the standard astrophysical
observations \citep{maggiore}. Direct measurements of the amplitude of this
background therefore are of fundamental importance for understanding the
evolution of the universe when it was younger than one minute. Here we
report direct limits on the amplitude of the stochastic gravitational-wave
background using the data from a two-year science run of the Laser
Interferometer Gravitational-wave Observatory (LIGO) \citep{S1}. Our result
constrains the energy density of the stochastic gravitational-wave
background normalized by the critical energy density of the universe, in the
frequency band around 100 Hz, to be less than $6.9\times 10^{-6}$ at 95\% confidence. The data
rule out models of early universe evolution with relatively large
equation-of-state parameter \citep{boylebuonanno}, as well as cosmic
(super)string models with relatively small string tension \citep{cspaper} that
are favoured in some string theory models \citep{cstrings}. This search for the stochastic
background improves upon the indirect limits from the Big Bang
Nucleosynthesis \citep{maggiore,BBN} and cosmic microwave background
\citep{smith} at 100 Hz.}

According to the general theory of relativity, gravitational waves (GWs) are
produced by accelerating mass distributions with a quadrupole (or higher) moment. Moreover,
in the early phases of the evolution of the universe, they can be produced by the mechanism of
amplification of vacuum fluctuations. Once produced, GWs
travel through space-time at the speed of light, and are essentially unaffected by the matter
they encounter. As a result, GWs emitted shortly after the Big Bang (and observed today)
would carry unaltered information about the physical processes that generated them.
These waves are expected to be generated by a large number of
unresolved sources, forming a stochastic gravitational-wave background (SGWB) that is usually described in terms
of the GW spectrum:
\begin{equation}
\Omega_{\rm GW}(f) = \frac{f}{\rho_c} \; \frac{d \rho_{\rm GW}}{df}\,,
\end{equation}
where $d\rho_{\rm GW}$ is the energy density of gravitational
radiation contained in the frequency range $f$ to $f+df$
and $\rho_c$ is the critical energy density of
the universe \citep{allenromano}. Many cosmological mechanisms for
generation of the SGWB exist, such as the inflationary models
\citep{PA1,PA2}, pre-big-bang models \citep{pbb,BMU,MB},
electroweak phase transition \citep{PT2}, and cosmic strings \citep{kibble76,cstrings,CS3,
cspaper}. There are also astrophysical mechanisms, such
as due to magnetars \cite{magnetars} or rotating neutron stars \citep{RFP}.
\begin{figure}
\includegraphics[width=3.5in]{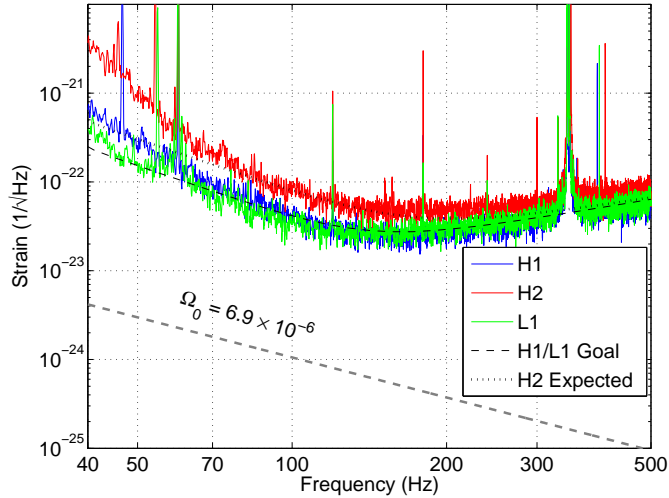}
\caption{{\bf Sensitivities of LIGO interferometers.}
LIGO interferometers reached their design sensitivity in November
2005, resulting in the
interferometer strain noise at the level of $3\times 10^{-22}$ rms in
a 100 Hz band around 100 Hz. This figure shows typical strain sensitivities
of LIGO interferometers during the subsequent science run S5.
Also shown is the strain amplitude corresponding
to the upper limit on the GW energy density presented in this paper
(gray dashed line). Note that this upper limit is $\sim 100$ times lower
than the individual interferometer sensitivities, which illustrates
the advantage of using the cross-correlation technique in this analysis.}
\label{fig_strain}
\end{figure}

The physical manifestation of GWs consists
of stretching and compressing the spatial dimensions orthogonal to
the direction of wave propagation, producing strain in an oscillating
quadrupolar pattern. A Michelson interferometer with suspended mirrors
\citep{S1} is well suited to measure this differential
strain signal due to GWs. Over the past decade, LIGO has
built three such multi-kilometer interferometers, at two locations
\cite{S1}: H1 (4~km) and H2 (2~km) share the same facility
at Hanford, WA, and L1 (4~km) is located in Livingston Parish, LA.
LIGO, together with the 3~km interferometer Virgo \cite{Virgo} in
Italy and GEO \cite{GEO} in Germany, forms a network of GW observatories.
LIGO has completed the science
run S5 (between November 5, 2005 and September 30, 2007), acquiring one year of data coincident among H1, H2 and
L1, at the interferometer design sensitivities (Fig. \ref{fig_strain}).

The search for the SGWB using LIGO data is performed by cross-correlating strain data
from pairs of interferometers \citep{allenromano}. In the frequency domain, the cross-correlation between
two interferometers is multiplied by a filter function $\tilde{Q}(f)$ (c.f. Data Analysis Supplement):
\begin{equation}
\label{optfilt1} \tilde{Q}(f) = \mathcal{N} \; \frac{\gamma(f)
\Omega_{\rm GW}(f) H_0^2}{f^3 P_1(f) P_2(f)} \; .
\end{equation}

This filter optimizes the signal-to-noise ratio,
enhancing the frequencies at which the signal of the
template spectrum $\Omega_{\rm GW}(f)$ is strong, while suppressing the frequencies at which
the detector noise ($P_1(f)$ and $P_2(f)$) is large. In Eq. \ref{optfilt1}, and throughout
this letter, we assume the present value of the Hubble parameter
$H_0 = 72$ km/s/Mpc \cite{hubble}, and use
$\gamma(f)$ to denote the overlap reduction function \cite{allenromano}, arising from
the overlap of antenna patterns of interferometers at different
locations and with different orientations. For the H1-L1 and H2-L1
pairs the sensitivity above roughly 50 Hz is attenuated due to the overlap reduction.
Since most theoretical models in the LIGO frequency band are characterized
by a power law spectrum, we assume a power law template GW spectrum
with index $\alpha$: $\Omega_{\rm GW}(f) = \Omega_{\alpha} ( f/100 {\rm \;
Hz})^{\alpha}$. The normalization constant $\mathcal{N}$ in Eq.
\ref{optfilt1} is chosen such that the expected value of the optimally filtered cross-correlation
is $\Omega_{\alpha}$.
\begin{figure}
\includegraphics[width=3.5in]{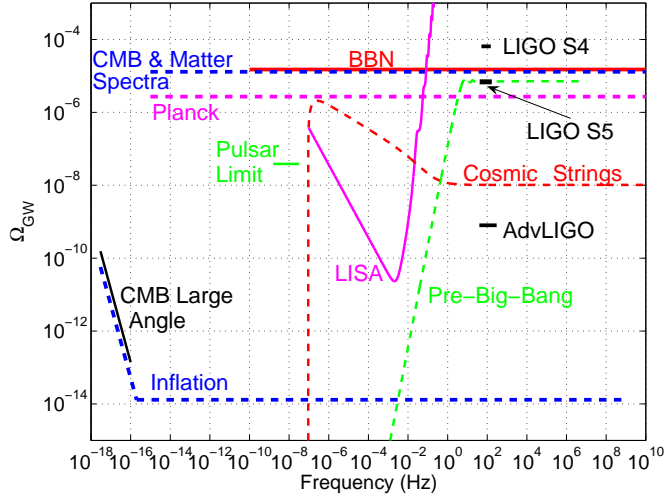}
\caption{{\bf Comparison of different SGWB measurements and models.} 
The 95\% upper limit presented here, $\Omega_0 < 6.9 \times 10^{-6}$ (LIGO S5), applies in the frequency band 41.5-169.25 Hz, and is compared to the previous LIGO S4 result \cite{S4paper} and to the projected Advanced LIGO sensitivity \cite{aligo1}. Note that the corresponding S5 95\% upper bound on the total gravitational-wave energy density in this band, assuming frequency independent spectrum, is $9.7 \times 10^{-6}$. The indirect bound due to BBN \cite{maggiore,BBN} applies to $\Omega_{\rm BBN} = \int \Omega_{\rm GW} (f) d(\ln f)$ (and not to the density $\Omega_{\rm GW} (f)$) over the frequency band denoted by the corresponding horizontal line, as defined in Equation \ref{BBNeq}. A similar integral bound (over the range 
$10^{-15}$ - $10^{10}$ Hz) can be placed using CMB and matter power spectra \citep{smith}. Projected
sensitivities of the satellite-based Planck CMB experiment \citep{smith} and LISA GW detector \citep{LISA} are also shown.
The pulsar bound \cite{pulsar} is based
on the fluctuations in the pulse arrival times of millisecond pulsars and applies at frequencies around $10^{-8}$ Hz. Measurements of the CMB at
large angular scales constrain the possible redshift of CMB photons due to the SGWB, and therefore
limit the amplitude of the SGWB at largest wavelengths (smallest frequencies) \cite{BBN}. Examples of
inflationary \citep{PA1,PA2}, cosmic strings \citep{kibble76,cstrings,CS3,cspaper},
and pre-big-bang \citep{pbb,BMU,MB} models are also shown (the amplitude and the spectral shape
in these models can vary significantly as a function of model parameters).}
\label{landscape}
\end{figure}

We apply the above search technique to the data acquired by LIGO during
the science run S5. We include two interferometer pairs: H1-L1 and H2-L1.
Summing up the contributions to the cross-correlation in
the frequency band 41.5-169.25 Hz, which contains 99\% of
the sensitivity, leads to the
final point estimate for the frequency independent GW spectrum ($\alpha=0$):
$\Omega_0 = (2.1 \pm 2.7) \times 10^{-6}$, where the quoted
error is statistical. We calculate the Bayesian 95\% confidence upper limit for
$\Omega_0$,
using the previous LIGO result (S4 run \cite{S4paper}) as a prior for $\Omega_0$ and
averaging over the interferometer calibration uncertainty. This procedure yields the 95\% confidence
upper limit $\Omega_0 < 6.9 \times 10^{-6}$. For other values of the power index $\alpha$ in the
range between $-3$ and $3$, the 95\% upper limit varies between $1.9\times 10^{-6}$ and
$7.1\times 10^{-6}$. These results constitute more than an order of magnitude
improvement over the previous LIGO result in this frequency region
\cite{S4paper}. Fig. \ref{landscape} shows this result in comparison with other observational
constraints and some of the cosmological SGWB models.
\begin{figure}
\includegraphics[width=3.5in]{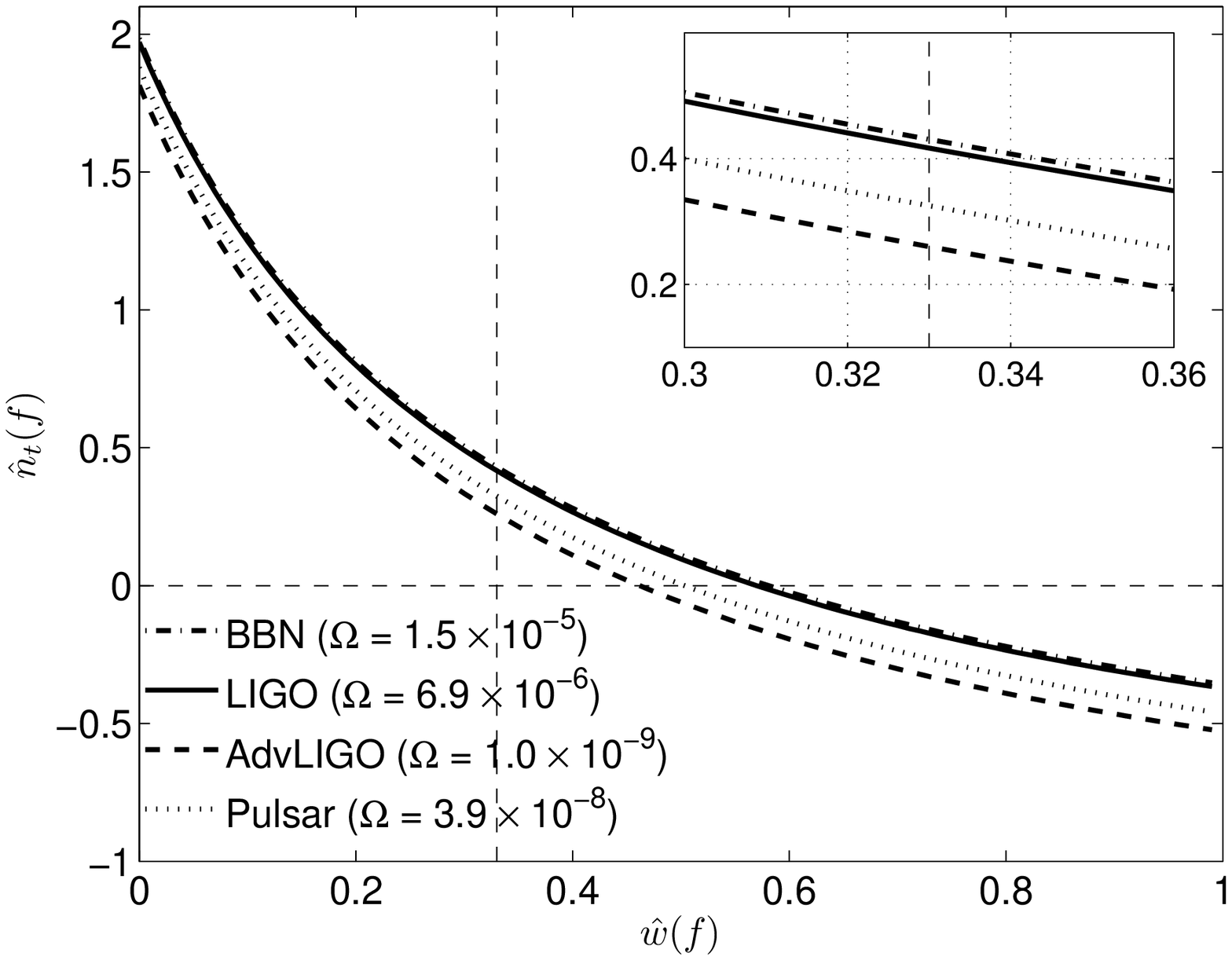}
\caption{{\bf Constraining early universe evolution.} The GW spectrum $\Omega_{\rm GW}(f)$ is related to the
parameters that govern the evolution of the universe \citep{boylebuonanno}:\\
$\Omega_{\rm GW}(f) = A \; f^{\hat{\alpha}(f)} \; f^{\hat{n}_t(f)} \; r$, where
$\hat{\alpha}(f) = 2 \; \frac{3\hat{w}(f) - 1}{3\hat{w}(f)+1}$,
$r$ is the ratio of tensor and scalar perturbation amplitudes
(measured by the cosmic microwave background (CMB) experiments), $\hat{n}_t(f)$
and $\hat{w}(f)$ are effective (average) tensor tilt and equation of state parameters
respectively, and $A$ is a constant depending on various cosmological parameters.
Hence, the measurements of $\Omega_{\rm GW}$ and $r$ can be
used to place constraints in the $\hat{w}-\hat{n}_t$ plane, {\it independently}
of the cosmological model. The figure shows the $\hat{w}-\hat{n}_t$ plane for $r=0.1$.
The regions excluded by the BBN
\cite{cyburt}, LIGO, and pulsar \cite{pulsar} bounds are above the corresponding curves
(the inset shows a zoom-in on the central part of the figure).
The BBN curve was calculated in \cite{boylebuonanno}. We note that the CMB bound \citep{smith}
almost exactly overlaps with the BBN bound.
Also shown is the expected reach of Advanced LIGO \cite{aligo1}. Note that these
bounds apply to different frequency bands, so their direct comparison is meaningful
only if $\hat{n}_t(f)$ and $\hat{w}(f)$ are frequency independent.
We note that for the simplest single-field inflationary model that still agrees with the
cosmological data, with potential $V(\phi) = m^2\phi^2/2$ (where $\phi$ is a scalar field of mass $m$), $r=0.14$ and $n_t(100 {\rm \; Hz})
= -0.035$ \cite{wmap}, implying a LIGO bound on the equation-of-state parameter of
$\hat{w}(100 {\rm \; Hz}) < 0.59$.}
\label{wvsn}
\end{figure}

Prior to the result described here, the most constraining bounds on the SGWB
in the frequency band around 100 Hz came from the Big-Bang-Nucleosynthesis (BBN)
and from cosmic microwave background (CMB) measurements.
The BBN bound is derived from the fact that
a large GW energy density at the time of BBN would
alter the abundances of the light nuclei produced in the process.
Hence, the BBN model and observations constrain
the total GW energy density at the time of nucleosynthesis
\cite{maggiore,BBN}:
\begin{equation}
\Omega_{\rm BBN} = \int \Omega_{\rm GW}( f ) \; d( \ln f ) < 1.1 \times 10^{-5} \; (N_{\nu} - 3),
\label{BBNeq}
\end{equation}
where $N_{\nu}$	(the {\it effective} number of neutrino species at the time of
BBN) captures the uncertainty in the radiation content during
BBN. Measurements of the light-element abundances, combined
with the Wilkinson Microwave Anistropy Probe (WMAP) data
give the upper bound $N_{\nu}-3 < 1.4$ \cite{cyburt}.
Similarly, a large GW background at the time of decoupling of CMB would alter
the observed CMB and matter power spectra. Assuming homogeneous initial
conditions, the total GW energy density at the time of CMB decoupling is
constrained to $\int \Omega_{\rm GW}( f ) \; d( \ln f ) < 1.3 \times 10^{-5} $
\citep{smith}. In the LIGO frequency band and for $\alpha=0$, these bounds
become: $\Omega_0^{\rm BBN} < 1.1 \times 10^{-5}$ and
$\Omega_0^{\rm CMB} < 9.5 \times 10^{-6}$. Our result
has now surpassed these bounds, which is one of the major
milestones that LIGO was designed to achieve. Moreover, the BBN and CMB bounds apply
only to backgrounds generated prior to the BBN and the CMB decoupling respectively,
while the LIGO bound also probes the SGWB produced later (this is the case, for example,
in cosmic strings models).

Our result also constrains models of the early universe evolution.
While the evolution of the universe following the BBN is well
understood, there is little observational data probing the
evolution prior to BBN, when the universe was less than one
minute old. The GW spectrum $\Omega_{\rm GW}(f)$ carries information about
exactly this epoch in the evolution. In particular, measuring $\Omega_{\rm GW}(f)$
is the best way to test for existence of presently unknown ``stiff'' energy
components in the early universe \cite{boylebuonanno},
for which a small density variation is associated with a large pressure change,
%(with the "equation of state" parameter larger than 1/3),
which could carry information about the physics of the inflationary era
\cite{stiffw1}.
Fig. \ref{wvsn} demonstrates how the result presented here can be used to constrain the
existence of these new energy components.
\begin{figure}
\includegraphics[width=3.5in]{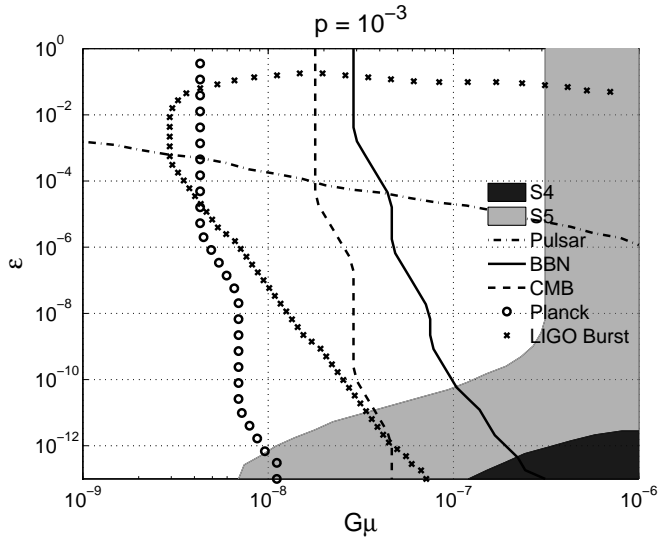}
\caption{{\bf Cosmic strings models.} The network of cosmic strings is usually parametrized by the string
tension $\mu$ (multiplied by the Newton constant $G$), and reconnection
probability $p$. The CMB observations limit $G\mu<10^{-6}$. If the size of the cosmic string loops is
determined by the gravitational back-reaction \cite{gbrloops1}, the size of
the loop can be parametrized by a parameter $\epsilon$ \cite{CS3} which is
essentially unconstrained.
The mechanism for production of GWs relies on cosmic string cusps: regions of
string that move at speeds close to the speed of light. If the cusp motion
points toward Earth, a detectable burst of gravitational radiation may be
produced \citep{CS3,SCMMCR}. The superposition of GWs from
all string cusps in the cosmic string network would produce a SGWB
\cite{cspaper}. This figure shows how different experiments probe the
$\epsilon - G\mu$ plane for a typical value of $p=10^{-3}$ \cite{cspaper} ($p$ is expected
to be in the range $10^{-4} - 1$). The excluded regions (always to
the right of the corresponding curves) correspond
to the S4 LIGO result \cite{S4paper}, this result, BBN bound \cite{BBN,cyburt}, CMB bound \citep{smith},
and the pulsar limit \cite{pulsar}. In particular, the bound presented in this paper excludes a new
region in this plane ($7\times 10^{-9} < G\mu < 1.5\times 10^{-7}$
and $\epsilon< 8\times 10^{-11}$), which is not accessible to any of the other measurements.
Also shown is the expected sensitivity for the search
for individual bursts from cosmic string cusps with LIGO S5 data
\cite{SCMMCR}. The region to the right of this curve is expected to produce at least one cosmic string
burst event detectable by LIGO during the S5 run. Note that this search is complementary to the search for the SGWB as it probes a
different part of the parameter space. Also shown is the region that will be probed by the
Planck satellite measurements of the CMB \citep{smith}.
The entire plane shown here will be accessible
to Advanced LIGO \cite{aligo1} SGWB search.}
\label{epsvsGmu}
\end{figure}

Our result also constrains models of
cosmic (super)strings. Cosmic strings were originally proposed as
topological defects formed during phase
transitions in the early universe \cite{kibble76}. More recently, it was
realized that fundamental strings may also be expanded to cosmological scales
\cite{cstrings}. Hence, searching for cosmic strings may provide a unique and powerful
window into string theory and into particle physics at the highest energy
scales. Fig. \ref{epsvsGmu} shows that our result,
along with other observations, can be used to constrain the parameters in the cosmic
string models. While our result is currently excluding a fraction of the
allowed parameter space, Advanced LIGO \cite{aligo1} is expected to
probe most of these models.
\begin{figure}
\includegraphics[width=3.5in]{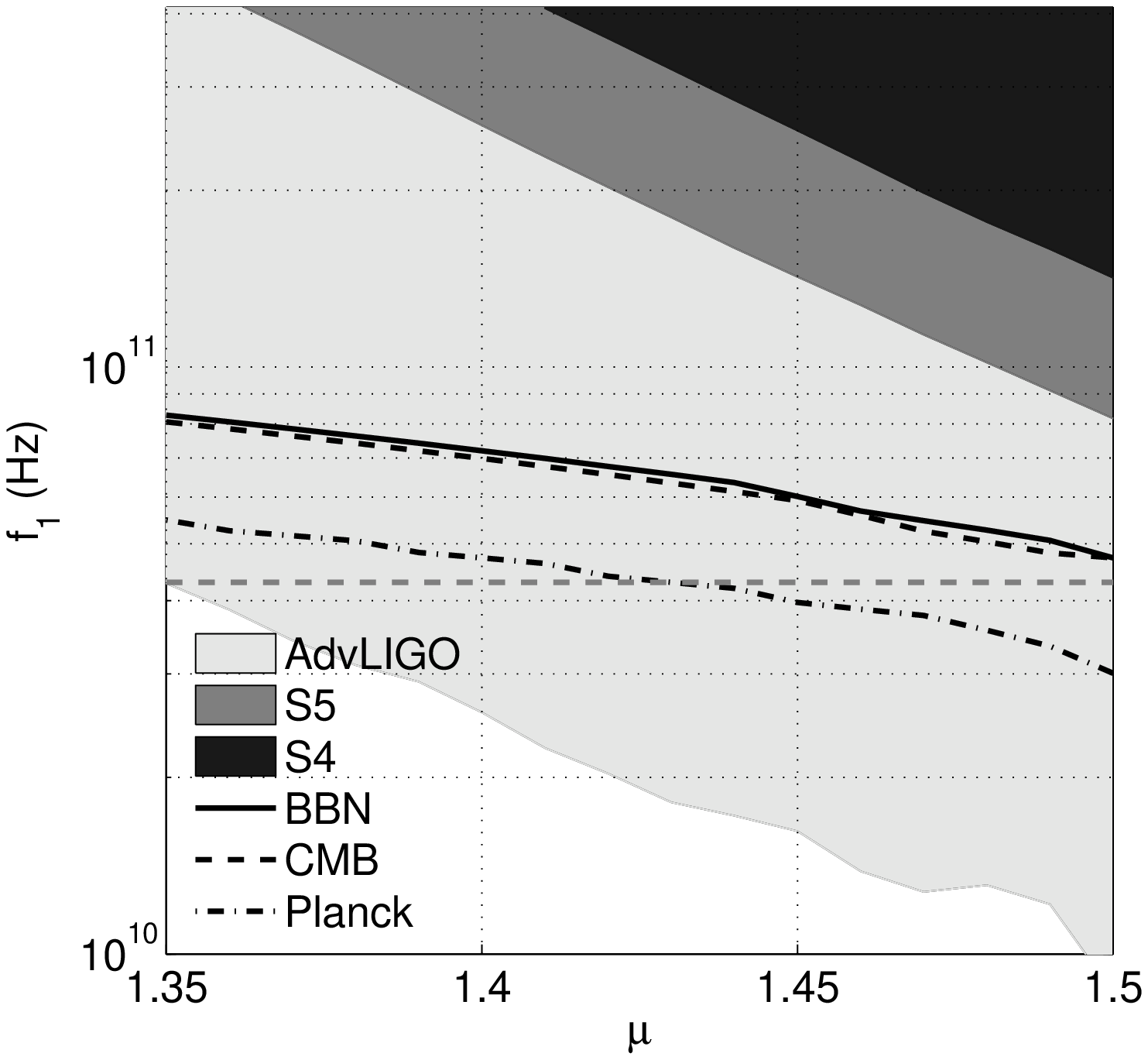}
\caption{{\bf Pre-Big-Bang models.} In the pre-Big-Bang model, the GWs are produced via the mechanism of
amplification of vacuum fluctuations, analogously to the standard inflationary model.
The typical GW spectrum increases as $f^3$ up to a turn-over
frequency $f_s$, above which $\Omega_{\rm GW}(f) \sim f^{3-2\mu}$ with
$\mu<1.5$. The spectrum cuts off at a frequency $f_1$, which is theoretically expected
to be within a factor of 10 from $4.3\times 10^{10}$ Hz (dashed horizontal line). This figure shows the
$f_1-\mu$ plane for a representative value of $f_s = 30$ Hz.
Excluded regions corresponding to the S4 result and to the result presented here
are shaded.
The regions excluded by the BBN \citep{BBN,cyburt} and the CMB \citep{smith} bounds are above the corresponding curves.
The expected reaches of the Advanced LIGO \citep{aligo1} and of the Planck satellite \citep{smith} are also shown.}
\label{f1vsmu}
\end{figure}

Measurements of the SGWB also offer the possibility of probing alternative models of
the early universe cosmology. For example, in the pre-Big-Bang model \citep{pbb,BMU,MB}
the universe starts off large and then undergoes a
period of inflation driven by the kinetic energy of a dilaton field, after which the
standard cosmology follows. Although more speculative than the standard cosmology model,
the pre-Big-Bang model makes testable predictions of the GW spectrum. As shown in Fig.
\ref{f1vsmu}, the BBN and CMB bounds are currently the most constraining for this model and
Advanced LIGO \cite{aligo1} is expected to surpass them.

\section{Data Analysis Supplement}

\subsection{Method}
Gravitational waves stretch and compress the spatial dimensions perpendicular to the direction
of wave propagation. In a Michelson interferometer with suspended mirrors, the gravitational wave would
cause stretching and shrinking of orthogonal arms, as shown in Figure \ref{gw}, which would result
in corresponding fluctuations in the laser intensity at the output of the interferometer. Hence, transient
or periodic gravitational waves would cause transient or periodic fluctuations in the output laser power.
A stochastic gravitational-wave background (SGWB) signal would cause random fluctuations in output laser power, which are
indistiguishable from various instrumental noise sources.
We hence search for a SGWB by cross-correlating strain data from pairs of interferometers, as described
in [8]. In particular, we define the following cross-correlation estimator:
\begin{figure}
\includegraphics[width=3.3in]{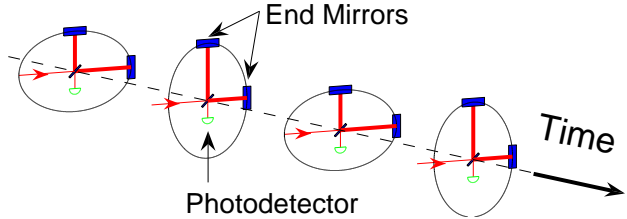}
\caption{{\bf Effect of a gravitational wave on an interferometer.}
A gravitational wave traveling perpendicular
to the plane of the interferometer stretches and compresses interferometer
arms in an alternating manner. The laser beam (entering from the left) is split
equally between the arms, the two new beams travel to and reflect back from the end mirrors,
and are superposed at the photo-detector (at the bottom). Changes in the arm lengths
cause the two beams to acquire different phases while traveling in the arms,
the differential component of which is observed as modulations in the laser
light intensity at the photo-detector.}
\label{gw}
\end{figure}
\begin{eqnarray}
Y & = & \int_{0}^{+\infty } df \; Y(f) \\
& = & \int_{-\infty }^{+\infty } df \int_{-\infty }^{+\infty } df'
\; \delta_T (f-f') \; \tilde{s}_1(f)^{*} \; \tilde{s}_2(f') \;
\tilde{Q}(f')\,, \nonumber \label{ptest}
\end{eqnarray}
where $\delta_T$ is a finite-time approximation to the Dirac delta
function, $\tilde{s}_1$ and $\tilde{s}_2$ are the Fourier
transforms of the strain time-series of two interferometers, and
$\tilde{Q}$ is a filter function. Assuming that the detector noise
is Gaussian, stationary, uncorrelated between the two
interferometers, and much larger than the GW signal, the variance
of the estimator $Y$ is given by:
\begin{eqnarray}
\sigma_Y^2 & = & \int_0^{+\infty} df \; \sigma_Y^2(f) \nonumber \\
& \approx & \frac{T}{2} \int_0^{+\infty} df P_1(f) P_2(f) \mid
\tilde{Q}(f) \mid^2\,, \label{sigma}
\end{eqnarray}
where $P_i(f)$ are the one-sided strain power spectral densities (PSDs)
of the two interferometers and $T$ is the measurement time.
Optimization of the signal-to-noise ratio leads to the following
form of the optimal filter [8]:
\begin{equation}
\label{optfilt2} \tilde{Q}(f) = \mathcal{N} \; \frac{\gamma(f)
\Omega_{\rm GW}(f) H_0^2}{f^3 P_1(f) P_2(f)} \; ,
\end{equation}
where $H_0$ is the present value of the Hubble parameter, assumed below to take the value
$H_0 = 72$ km/s/Mpc [19], and
$\gamma(f)$ is the overlap reduction function [8], arising from
the overlap of antenna patterns of interferometers at different
locations and with different orientations. For the Hanford-Livingston
pairs the sensitivity above 50 Hz is attenuated due to the overlap reduction,
while the identical antenna patterns of the colocated Hanford interferometers
imply $\gamma(f) = 1$. Hence, the colocated Hanford interferometer pair is
more sensitive to the isotropic SGWB than the Hanford-Livingston pairs, but it is
also more susceptible to environmental and instrumental correlations. For this reason,
this pair is not included in the analysis presented here.
%\begin{figure}
%\includegraphics[width=3.3in]{overlap_reduction.eps}
%\caption{Overlap reduction function for the Hanford-Hanford pair
%(black solid) and for the Hanford-Livingston pair (gray dashed).}
%\label{overlap}
%\end{figure}
%In Equations \ref{optfilt1} and \ref{optfilt2}, $S_{\rm GW}(f)$ is
%the strain power spectrum of the stochastic GW background to be
%searched.
Since most theoretical models in the LIGO frequency band are characterized
by a power law spectrum, we assume a power law template GW spectrum
with index $\alpha$,
\begin{equation}
\Omega_{\rm GW}(f) = \Omega_{\alpha} \Bigg( \frac{f}{100 {\rm \;
Hz}} \Bigg)^{\alpha}.
\end{equation}
The normalization constant $\mathcal{N}$ in Equation
\ref{optfilt2} is chosen such that $<Y> = \Omega_{\alpha}$.

\subsection{Results}
Our results are based on the LIGO data acquired during
the science run S5, which took place between November 5, 2005 and
September 30, 2007. Virgo [19] and GEO [20] detectors
were also operating during some parts of this science run. However,
due to their lower strain sensitivities around 100 Hz, these
interferometers were not included in the search presented here.
We analyzed the H1-L1 and H2-L1 interferometer pairs.
The data for each interferometer pair was divided into
60 sec segments, down-sampled to 1024 Hz, and high-pass filtered
with a $6^{\rm th}$ order Butterworth filter
with 32 Hz knee frequency. Each segment $I$ was Hann-windowed
and estimators $Y_I(f)$ and $\sigma_I(f)$ were evaluated with 0.25 Hz
resolution. To recover the loss
of signal-to-noise due to Hann-windowing, segments were 50\% overlapped.
A weighed average was performed over all segments from both
interferometer pairs, with inverse variances as weights.
\begin{figure}
\includegraphics[width=3.3in]{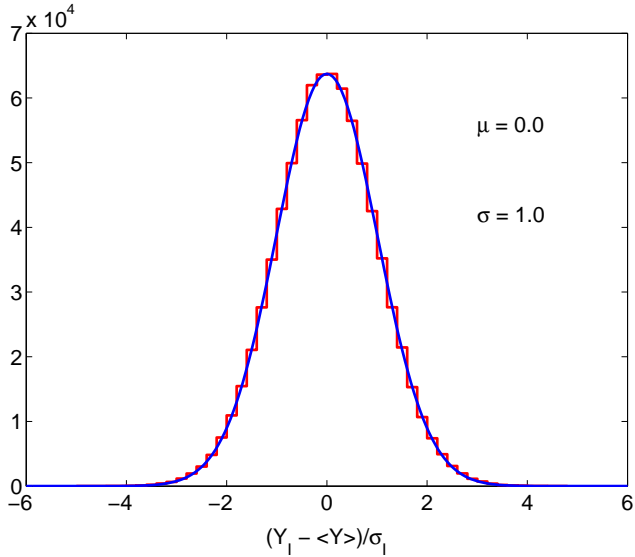}
\caption{Histogram of the fluctuations of the
estimator $Y_I$ over segments $I$ around the mean, normalized by the
standard deviation $\sigma_I$ is shown in red (for the H1-L1 pair).
The blue curve shows the Gaussian fit to the histogram, which has zero mean
and unit variance. The Kolmogorov-Smirnov test statistic (comparing the histogram
and the fit) is 0.2 for H1-L1 (0.4 for H2-L1), indicating that the data is
indeed Gaussian-distributed, and that the estimate of the theoretical
variance $\sigma^2_I$ is reliable.}
\label{resdistr}
\end{figure}

\begin{figure}
\includegraphics[width=3.3in]{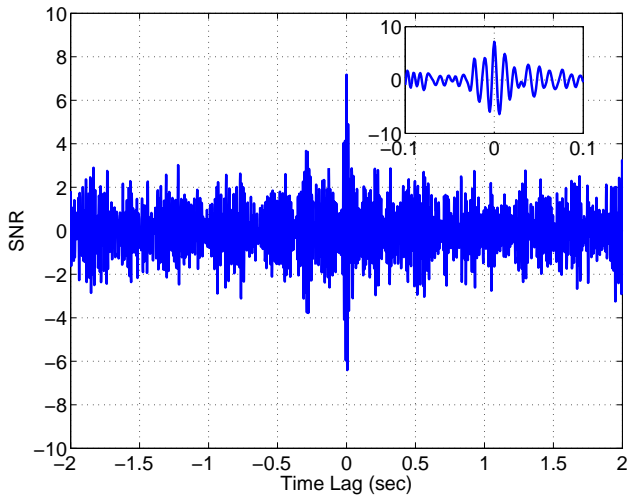}
\caption{Signal-to-noise ratio for the recovery of a software simulation with H1-L1
data with $\Omega_0^{simulated} = 3.8\times 10^{-5}$ is shown
as a function of the time-lag between the two interferometers. The inset
shows the zoom-in around zero-lag: the signal
is recovered well for zero-lag (${\rm SNR} \approx 7.2$), but it disappears quickly with time-lag of
$\pm 30$ ms.}
\label{inj2}
\end{figure}

The data were preselected to avoid digitizer saturation effects, periods
with unreliable calibration, and periods suffering from known instrumental
transient disturbances. In addition,
about 3\% of the segments were found to deviate from the assumption of
stationary noise: the difference between $\sigma_I$ and
$\sigma$ calculated using the neighboring segments exceeded 20\% for these
segments, and they were not included in the analysis. The 20\% threshold
is optimal as it yields gaussian distribution of the data (c.f. Figure
\ref{resdistr}), while minimizing the amount of eliminated data. The data quality
selection was performed blindly, using an un-physical 0.5-sec time-shift
between the two interferometers (a broadband SGWB covering the range of
$\sim 100$~Hz is expected to have coherence time $\sim$ 10~ms, as also
depicted in Figure \ref{inj2}). Once the data selection was completed, the final
zero-lag analysis was performed.
The selected segments amount to 292 days of exposure time for
H1-L1 (294 days for H2-L1).
\begin{figure}
\includegraphics[width=3.3in]{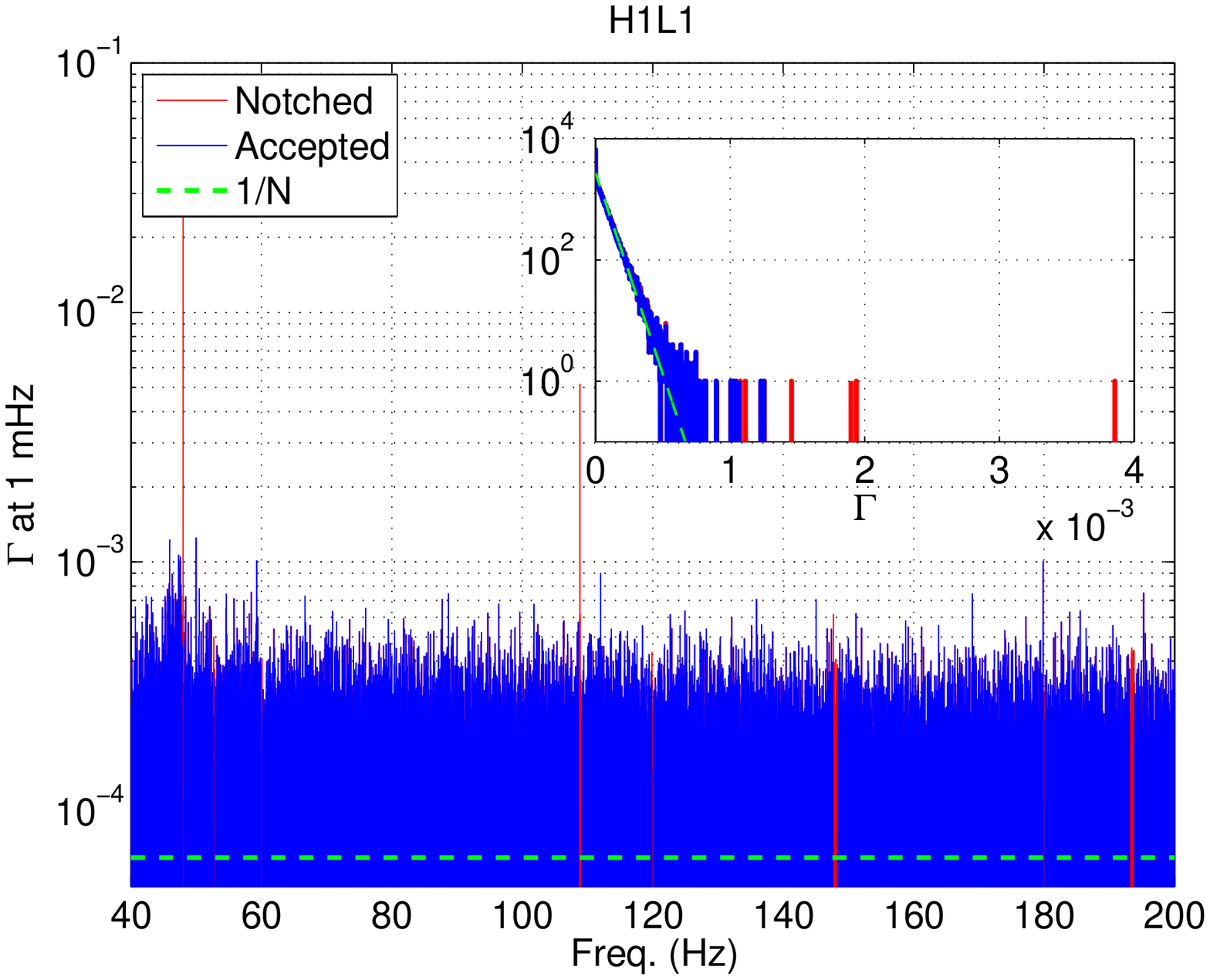}
\includegraphics[width=3.3in]{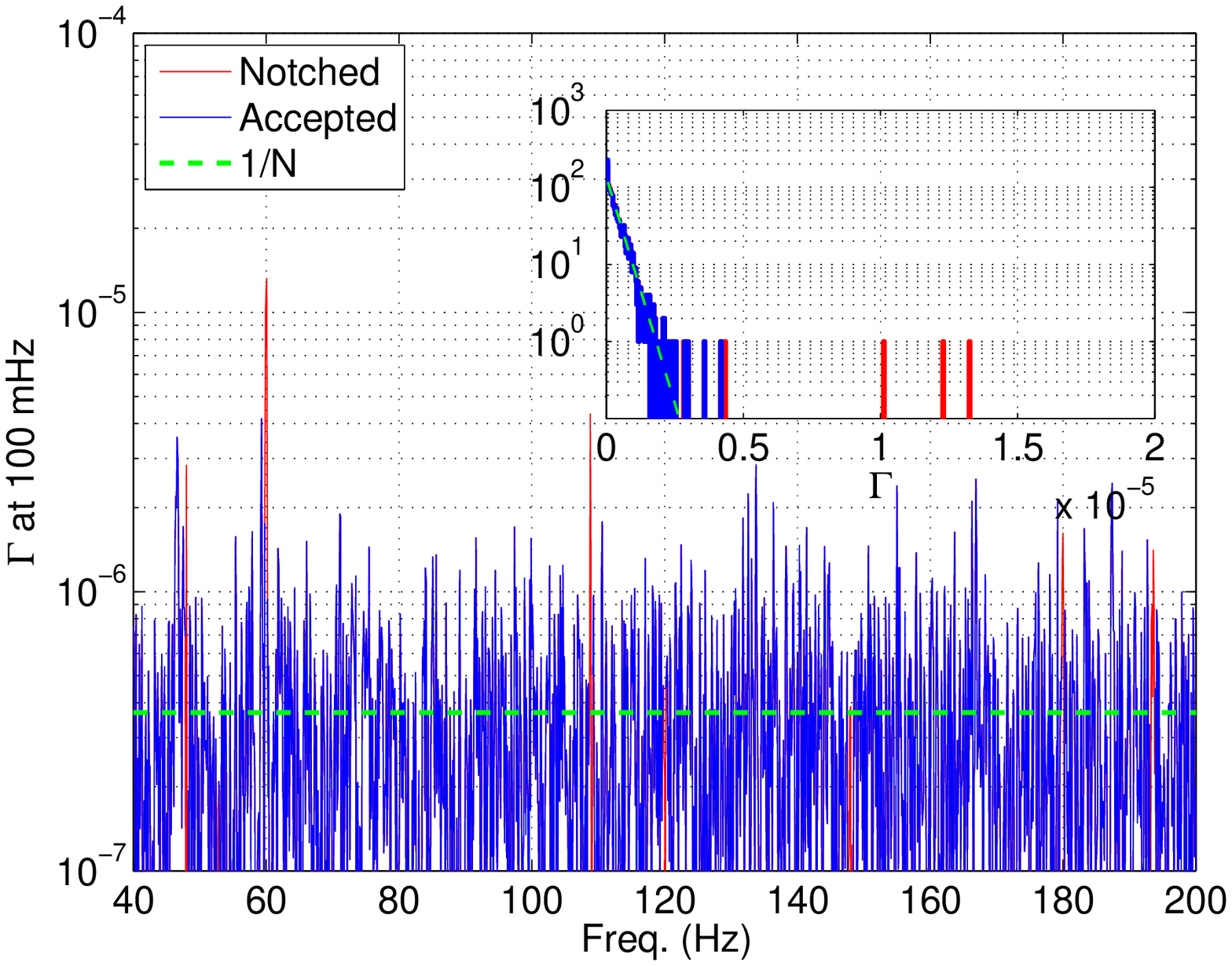}
\caption{Coherence between H1 and L1 strain data is shown at 1 mHz resolution
(top) and 100 mHz resolution (bottom). The insets show the histograms of the coherence
along with the expected exponential distribution. Note that after notching the
contaminated bins (red), the remaining frequencies follow the expected exponential
distribution. Note: $N$ denotes the number of averages used in the calculation.}
\label{coherence}
\end{figure}

To identify potentially contaminated frequency bins, we calculated the coherence
between H1 and L1 (and H2 and L1) over the entire S5 run. The coherence is defined as
\begin{equation}
\Gamma(f) = \frac{| \langle P_{12}(f) \rangle |^2}{\langle P_1(f) \rangle
\langle P_2(f) \rangle},
\end{equation}
where $\langle P_{12}(f) \rangle$ is the average strain cross-spectral density
between two interferometers and $\langle P_i(f) \rangle $ is the average strain
power-spectral density for the interferometer $i$. These
calculations have revealed several
instrumentally correlated lines between each pair of interferometers: 16 Hz harmonics
(associated with the data acquisition clock), 60 Hz harmonics (AC power line), and
injected simulated pulsar signals (52.75~Hz, 108.75~Hz, 148~Hz, 193.5~Hz, and 265.5~Hz).
These lines were found to be correlated between instruments in the blind analysis,
and were excluded from the final zero-lag analysis.
Figure \ref{coherence} shows the coherence between H1 and
L1 strain data at 1 mHz and 100 mHz resolutions.
\begin{figure}
\includegraphics[width=3.3in]{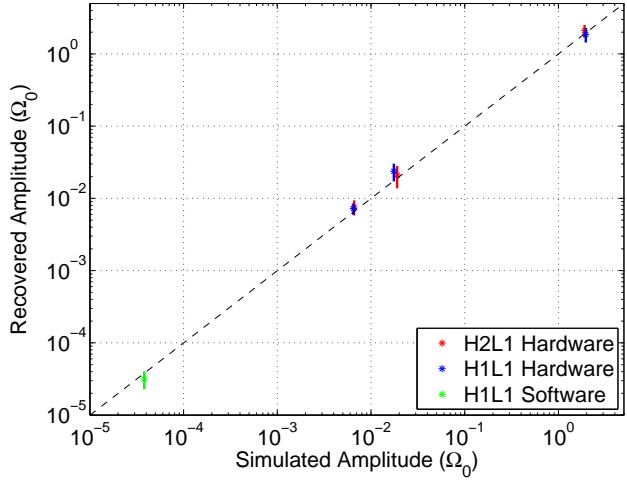}
\caption{Stochastic signal simulations in hardware for H1-L1 (blue) and H2-L1 (red),
and in software (H1-L1, green) are shown. The error bars denote $2\sigma$ ranges.}
\label{inj1}
\end{figure}

The search algorithm described here is
verified using signal simulations. The simulations are performed in
hardware (by physically moving the interferometer mirrors coherently
between interferometers), in which
case they are short in duration and strong in amplitude. They are also
performed in software, by adding the stochastic signal to the interferometer
data, in which case they can be long in duration and relatively weak in
amplitude. Three hardware simulations were performed, with amplitudes of
$\Omega_0 \approx 2$ (20~min long), $2\times 10^{-2}$ (20~min long),
and $6.5\times 10^{-3}$ ($\sim 3.8$~hours long) and they
were successfully recovered (within experimental uncertainties) for both H1-L1
and H2-L1 pairs. A software simulation was performed and successfully
recovered using about 1/2 of the H1-L1 data, with the amplitude of $\Omega_0 =
3.8\times 10^{-5}$. Figures \ref{inj1} and \ref{inj2} demonstrate the recovery
of both hardware and software simulations.
\begin{figure}
\includegraphics[width=3.3in]{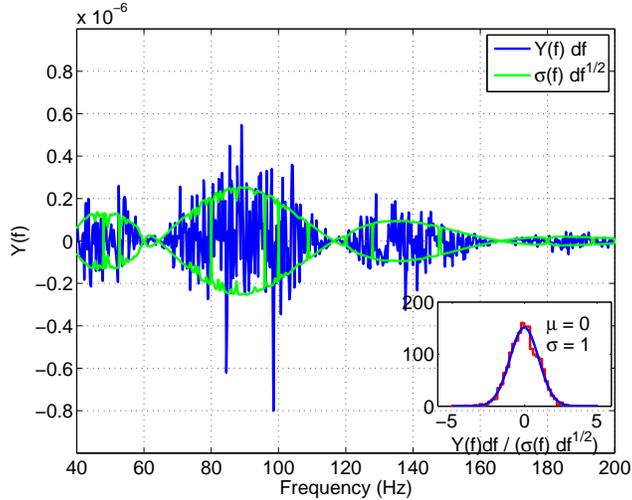}
\caption{$Y(f)$ and $\sigma(f)$ obtained by combining the H1-L1 and H2-L1
data from the S5 run. The inset shows that the ratio of the two spectra is
consistent with a Gaussian of zero mean and unit variance.}
\label{spectra}
\end{figure}
\begin{figure}
\includegraphics[width=3.3in]{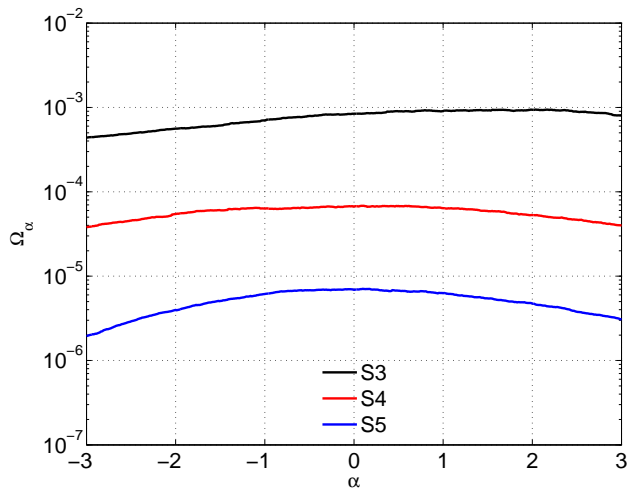}
\caption{Upper limit is shown as a function of the power index $\alpha$ for
several LIGO results: based on the previous runs S3 and S4
and the S5 result presented here.}
\label{ULvsalpha}
\end{figure}

We apply the above search technique to the data acquired by LIGO during
the science run S5, repeating it for the interferometer pairs H1-L1 and H2-L1.
We treat the data from the two pairs as uncorrelated, although H1 and H2 are known
to suffer from instrumental and environmental correlations. We have verified
that the level of the H1-H2 correlations is sufficiently small
that it could affect the result presented here by less than 1\%.
The resulting composite spectrum for the frequency independent template
($\alpha=0$) is shown in Figure \ref{spectra}.
Integrated over the frequency band 41.5-169.25 Hz, which contains 99\% of
the sensitivity as determined by the variance integrand, this leads to the
final point estimate for the frequency independent GW spectrum:
$\Omega_0 = (2.1 \pm 2.7) \times 10^{-6}$, where the quoted
error is statistical. We calculate the Bayesian posterior distribution
for $\Omega_0$ using this result. For the prior distribution of
$\Omega_0$ we use our previously published posterior distribution from the
earlier S4 run [22]. We also
marginalize over the calibration uncertainty, which is the dominant systematic
error in this analysis and was estimated to be 13.4\% for L1 and 10.3\% for H1 and H2.
With these assumptions, the final 95\% confidence upper limit is
$\Omega_0 < 6.9 \times 10^{-6}$. Figure \ref{ULvsalpha}
shows the 95\% confidence upper limit as a function of the power index $\alpha$
of the template spectrum. This result is more than an order of magnitude
improvement over the latest LIGO result in this frequency region [22].

\subsection{Outlook}
LIGO and Virgo are planning significant upgrades to their interferometers, known as Advanced LIGO and Advanced Virgo.
These upgrades will improve the interferometers' strain sensitivities by 10 times across the entire frequency band,
and they will extend the sensitive
frequency band down to $\sim 10$ Hz. Consequently, the network of advanced detectors will be able to probe
the isotropic SGWB at the level of $\Omega_{{\rm GW}} \sim 10^{-9}$ or better. Moreover, while searches for isotropic SGWB
tend to be dominated by pairs of nearby or co-located detectors, the presence of the third location in the network is crucial for
searches for non-isotropic SGWB. Hence, the network of advanced detectors is expected to produce detailed
maps of the gravitational-wave sky, potentially revealing non-isotropic sources of SGWB, such as point sources or
sources distributed in the galactic plane. Techniques for performing the searches for non-isotropic SGWB are currently
under development.

{\bf Author Contributions:} The LIGO Scientific Collaboration (LSC) and the Virgo Collaboration are organized into several working groups, each focusing on a different aspect of the experiment. Each author is associated with one or more of these groups. Moreover, all of the authors participated in the acquisition of the data that led to this letter. As described in the MoU between the LSC and Virgo, there are joint data analysis groups:
the Burst Search Group (chairs E. Katsavounidis, P. Shawhan and P.Hello) performs searches for transient signals; the Compact Binary Coalescences Search Group (chairs S. Fairhurst, A. Weinstein and F. Marion) performs searches for compact binary coalescence signals; the Continuous-Waves Search Group (chairs K. Riles, G. Woan and C. Palomba) performs searches for continuous periodic signals; the Stochastic Search Group (chairs S. Ballmer, V. Mandic and G. Cella) performs searches for stochastic signals and is responsible for the analysis that led to the result presented here and for writing of this letter. The Stochastic Review Committee (chairs W. Anderson and F. Ricci) has conducted an extensive internal review of the method and the results presented in this letter. Similar review committees exist for the Burst, Compact Binary Coalescences, and Continuous-Waves groups. The Data Analysis Council (chairs M.A. Papa and G. Guidi) coordinates different analysis efforts.
On the LIGO side the Detector Characterization Group (chair G. Gonzalez) performs studies of LIGO detector performance; Calibration Group (chairs K. Kawabe and X. Siemens) establishes the calibration of LIGO detectors.
On the Virgo side the Reconstruction group (coordinator L. Rolland) provides a reconstructed and calibrated strain measurement while the Noise Study group (coord. E. Cuoco) studies the noise present in the detector. The Commissioning group (coord. E. Calloni) establishes proper detector tuning. The Detector Group (coord. P. Rapagnani) organizes detector operations and upgrades.
The Director of the LIGO Laboratory is J. Marx, the Deputy Director is A. Lazzarini, and the LSC spokesperson is D. Reitze. The Virgo Spokesperson is F. Fidecaro.

{\bf Supplementary Information} is linked to the online version of the paper at www.nature.com/nature.

{\bf Acknowledgements:} The authors gratefully acknowledge the support of the United States
National Science Foundation for the construction and operation of the
LIGO Laboratory, the Science and Technology Facilities Council of the
United Kingdom, the Max-Planck-Society, and the State of
Niedersachsen/Germany for support of the construction and operation of
the GEO600 detector,
and the Italian Istituto Nazionale di Fisica Nucleare and the French
Centre National de la Recherche Scientifique for the construction and
operation of the Virgo detector. The authors also gratefully acknowledge the support
of the research by these agencies and by the Australian Research Council,
the Council of Scientific and Industrial Research of India, the Istituto
Nazionale di Fisica Nucleare of Italy, the Spanish Ministerio de
Educacion y Ciencia, the Conselleria d'Economia Hisenda i Innovacio of
the Govern de les Illes Balears, the Royal Society, the Scottish Funding
Council, the Scottish Universities Physics Alliance, The National Aeronautics
and Space Administration, the Carnegie Trust, the Leverhulme Trust, the David
and Lucile Packard Foundation, the Research Corporation, and the Alfred
P. Sloan Foundation. This document has been assigned LIGO Laboratory
document number LIGO-P080099-12-Z.

{\bf Author Contributions} are listed in Supplementary Information.

{\bf Author Information:} Reprints and permissions information is available at npg.nature.com/reprintsandpermissions.
Correspondence and requests for materials should be addressed to V. Mandic (mandic@physics.umn.edu).

\end{document}